\documentclass[usegraphicx,usedcolumn,usenatbib]{mn2e}
\def\ec{{$\eta$ Carinae}}

\begin{document}

\title
{G287.84--0.82: An Infrared Star Cluster in the Carina Nebula}
\author[G. H\"agele et al.]
{G. F. H\"agele$^{1}$\thanks{PhD fellow of CONICET, Argentina; ghagele@fcaglp.edu.ar},
J. F. Albacete Colombo$^{1}$\thanks{Post Doctoral fellow of CONICET, Argentina.},
R. H. Barb\'a$^{1}$$^{2}$\thanks{Member of Carrera del Investigador Cient\'{\i}fico, 
CONICET, Argentina.},
G. L. Bosch$^{1}$\thanks{Member of Carrera del Investigador Cient\'{\i}fico,
CONICET, Argentina.}
\\
$^{1}$ Facultad de Ciencias Astron\'omicas y Geof\'{\i}sicas de 
La Plata, Paseo del Bosque S/N, 1900 La Plata, Argentina.\\
$^{2}$ Departamento de F\'{\i}sica, Universidad de La Serena, Benavente 980, La Serena, Chile\\ }

\maketitle

\begin{abstract}
We have studied the properties of an infrared cluster embedded in the gas 
and dust of the southern part of the Carina Nebula (NGC\,3372), where the 
probable existence of current star formation has already been predicted. 
We used mid-infrared (A \& C bands) and near-infrared (J\,H\,K$_{\rm s}$) 
images from the MSX and the 2MASS surveys respectively, combined with an 
optical H$_\alpha$ narrow-band filter image obtained at the CTIO. 
The infrared star cluster has at least 17 members, and its parameters, 
radius and stellar density are in very good agreement with high- to 
intermediate-mass star formation scenarios. 
The detected IR sources have roughly the same intrinsic infrared excess
determined from their position in colour-colour and colour-magnitude diagrams, 
suggesting that these objects could be related to pre-main sequence stars of 
high to intermediate mass. Furthermore, we present a low-dispersion spectrum 
of LS\,1883 (O9.5\,V) star located near the centre of the IR cluster.
The position of this object in the colour-colour and colour-magnitude 
IR diagrams lies close to the reddening vector of an ZAMS O9\,V spectral type 
star, and it seems to be the first star of this cluster to emerge. 
All these facts are consistent with the current star-forming 
scenarios associated with highly embedded star clusters.

\end{abstract}


\section{Introduction}

The Carina Nebula (NGC\,3372) is one of the most interesting massive 
star-forming regions in the Galaxy which can be observed from the Southern 
hemisphere, containing a large population of hot stars 
\citep{1971ApJ...167L..31W,1973ApJ...179..517W}. 
This extraordinary region is available to observations using 
medium-sized telescopes and it hosts several well-known young open clusters, 
such as Trumpler 14, 15 and 16, Collinder 228 and 232, and Bochum 10 and 11, 
\citep{1993AJ....105..980M,1995RMxAC...2...51W}.

NGC\,3372 is an evolved H{\sc ii} region, with dust and neutral gas evacuated 
from its core by the action of strong winds from its hot massive stars. 
However, embedded star-forming cores suggest that star formation has not 
ceased \citep{2000ApJ...532L.145S}. 
These cores allow us to study the very initial phases of star 
formation which, in general, are deeply embedded in dust and/or are located 
in heavily reddened lines of sight. The near-infrared (near-IR) and 
mid-IR domains are therefore necessary to study them 
\citep{1991fesc.book....3L}.
Recently, \cite{2002MNRAS.331...85R} discovered several groups of 
diffuse IR sources and studied the interaction between the young 
stars and the molecular material that surrounds them. 
\cite{2004A+A...418..563R} found 12 candidate embedded clusters
using mid-IR Midcourse Space Experiment (MSX) data; some of these sources 
are also proposed as candidates by \cite{2000ApJ...532L.145S}.

Among all these candidates, G287.84--0.82 is a bright nebulosity near LS\,1883 
(CPD-59$^{\circ}$\,2661; $\alpha_{\rm 2000}$=10h\,45m\,53.6s, 
$\delta_{\rm 2000}$=-59$^{\circ}$\,57\arcmin\,02.2\arcsec), which
seems to be a prime candidate of current star formation.
\cite{2001A&A...376..434D} and \cite{2003A&A...400..533D} searched for 
candidates to star clusters and groups between $|b|$\,$<$\,10$^{\circ}$ and
230$^{\circ}$\,$<$\,{\it l}\,$<$\,350$^{\circ}$ using the 
Two-Micron All-Sky Survey (2MASS)\footnote{http://www.ipac.caltec.edu/2mass/}.
They have suggested that this region could be a partially resolved cluster in 
a reflection nebula in agreement with the previous classification vdBH-RN43 
\citep{1975AJ.....80..208V}. They have also given an estimate for the angular 
diameter of D=1.5\,arcmin at a distance of 2.5\,kpc 
(equivalent to a linear diameter of 1.1 pc), and a possible multiplicity of 
about 10 members. \cite{2003A&A...399..141M}, who merged several catalogues 
of reflection nebula in order to create a uniform catalogue containing 
913 objects, also classified it as an object of this type.

The aim of this paper is to determine the properties of this 
deeply embedded IR star cluster, by using mid-IR 8\,-\,21 $\mu$m 
images obtained with the MSX and the 2MASS IR atlas 
\citep{1997ilsn.proc...25S} in the J, H and K$_{\rm s}$ image bands.
Also, we have included new spectroscopic data of LS\,1883 and SS73\,24,
two bright stars which appear projected towards the nebulosity,
in order to determine the actual
association of these sources with the embedded cluster.

\section{Observations and data reduction}

The Carina Nebula was observed in 1996 in the mid-IR by the MSX satellite. 
The characteristics of its instruments are given by 
\cite{1998ApJ...494L.199E}. 
We retrieved MSX A-band (8.28\,$\mu$m), C-band (12.13\,$\mu$m), 
D-band (14.65\,$\mu$m) and 
E-band (21.3\,$\mu$m) images from the IPAC 
database\footnote{http://irsa.ipac.caltech.edu/msx.html}.

We also obtained narrow-band [O{\sc iii}] 5007\,\AA, H$\alpha$ and [S{\sc ii}] 
6730\,\AA\,\,emission-line images of a large field
in Carina, centered at $\eta$Car, observed in 1999 May using a 
SITe 2048$\times$2048 CCD (24 $\mu$m pixels) mounted on the 
Curtis-Schmidt telescope at Cerro Tololo Inter-American Observatory (CTIO) 
with a scale of 2.32 arsec\,pixel$^{-1}$.

\begin{figure*}
\centering
\vspace{14.5cm}
\includegraphics{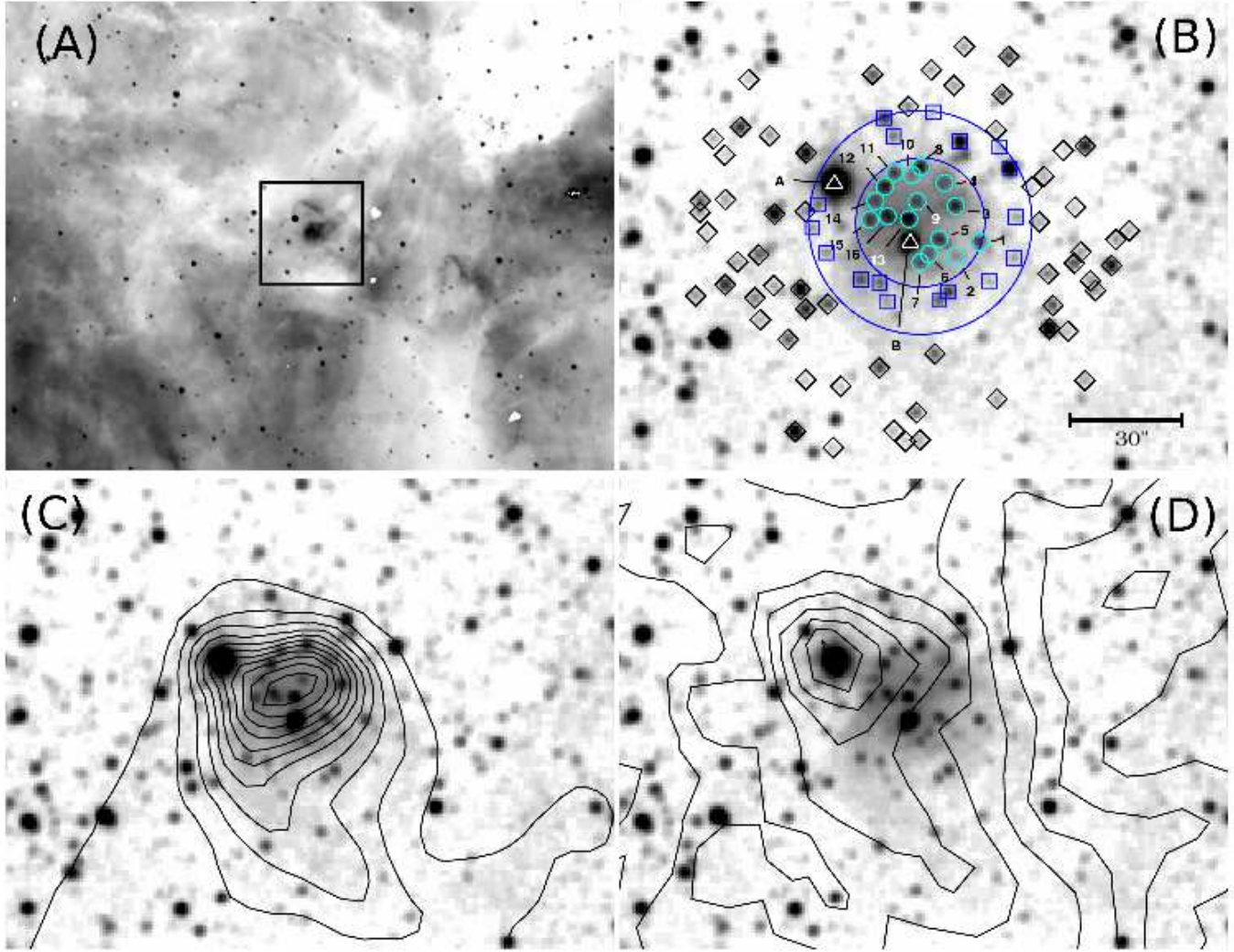}
\caption[]{Panel A: H$_\alpha$ image centered in 
$\alpha_{\rm 2000}$=10h\,45m\,53.6s and $\delta_{\rm 2000}$=-59$^{\circ}$\,57'\,02.0'' 
obtained at CTIO with the Curtis-Schmidt camera.
The box region shows the field of view of 160\,arcsec$\,\times\,$160\,arcsec of the 
2MASS H-band images presented in panels B, C and D.
Panel B: The members of the IR cluster labelled with small open  
circles are inside 22.5\,arcsec (first big circle); 
suspected foreground stars with small open squares are between
22.5 and 40\,arcsec (second big circle).
Diamonds mark the field stars (outside 
the second big circle). The stars labelled with A and B  
and marked with triangles are SS73\,24 and LS\,1883, respectively. 
Panel C shows the dust density distribution, 
traced by the single mid-IR (band A, $8\mu$m) emission, 
whereas panel D shows the contours of the intensity ratio of the A and C ($12.13\mu$m)
bands which traces dust temperature. It can be readily seen that, 
although the dust distribution is centred on the core of the IR 
cluster, SS73\,24 (source A) contributes with intrinsic warm dust emission.}
\label{cumulos}
\end{figure*}

The near-IR data retrieved from the 2MASS consist of a set 
of J\,(1.25\,$\mu$m), H\,(1.65\,$\mu$m) and K$_{\rm s}$\,(2.17\,$\mu$m) 
images centred on 
$\alpha_{\rm 2000}$=10h\,45m\,53.6s, 
$\delta_{\rm 2000}$=-59$^{\circ}$\,57\arcmin\,02\,arcsec, each about 300\,arcsec 
wide. These images have a spatial resolution of about 2\,arcsec.
We also retrieved photometric 
measurements of the same area from the 2MASS All-Sky Point Source Catalog 
\citep{2000IPAC} using the GATOR interface at the IPAC web site. 
However, we have limited our analysis to those sources whose errors 
are measurable (see table \ref{sources}) and whose quality parameters
guarantee that the reliability of the photometry is at least within the 
95 per cent confidence level (see the ``User's Guide to the 2MASS All-Sky 
Data Release'', also available from the IPAC site, for an explanation of the 
different types of 2MASS photometric errors).
These selection criteria removed roughly one-third of the 2MASS point sources 
available in the region. The completeness limits for our data set are 
about 15.5, 15.0 and 14.0 mag for J, H and K$_{\rm s}$ bands, 
respectively. We have only kept stars that are flagged for not having crowding
effects when determining their background, as this effect 
only slightly degrades the quality of the photometry, but does not introduce 
systematic effects on the magnitude determination. Note that 
source 4 has only been considered in order to determine the density, 
and it has not been used to estimate the global properties of the IR cluster 
(see discussion in Section 3). 

Low resolution spectra of LS\,1883 and SS73\,24 were obtained at Complejo
Astron\'omico El Leoncito (CASLEO)\footnote{Casleo is operated under 
agreement between CONICET and the National Universities of La Plata, 
C\'ordoba and San Juan} on 2001 June 30 with the REOSC
\footnote{Spectrograph 
Echelle Li\`ege (jointly built by REOSC and Li\`ege Observatory, 
and on long term loan from the latter)}
spectrograph attached to the Jorge Sahade 2.15-m telescope. We used the 
single dispersion mode which yields a spectral resolution of 
$\sim2.2$\,\AA\,pixel$^{-1}$. The background subtraction in both spectra was
performed in a very narrow window close to the star to minimize the
nebular contribution in H and He{\sc i} stellar line absorptions.
Moreover, in the case of LS\,1883, it was almost impossible to neatly 
subtract the nebular emission due to the clumpy aspect of the nebula around
the star.

The optical images were processed and analysed with 
IRAF\footnote{Image Reduction and Analysis Facility, distributed by NOAO, 
operated by AURA, Inc., under agreement with NSF.} routines in the usual 
manner.

\section{Analysis}

\subsection{Imaging}

Previous studies have discussed the relationship between the 
optical and MSX images of the dust pillar surrounding the source 
G28784--0.82. It is the brightest point source in the mid-IR and 
843\,MHZ radio continuum images in the southern part of the Carina nebula
\citep{2004A+A...418..563R}. Also, these authors proposed that the region
is a compact H{\sc ii} region containing at least a massive star surrounded 
by a cluster of lower-mass embedded stars. 
The inspection of the optical emission images shows that the bright nebular 
clump around LS\,1883 has dusty lanes crossing it, showing that the dust is 
intimately associated with the compact H{\sc ii} region. LS\,1883 was
classified as an O9.5\,V star by \cite{1984A+A...138..380W}, and we confirm 
this classification (see below). This star is placed slightly off 
centre to the south-east of the nebular clump, as seen in the 
emission line optical images, but in the IR images (see Figure 1) the
star is away from the centre of the cluster to the south. 
The optical images show a clear gradient in brightness and ionization, peaking
in nebulosity just around LS\,1883 and decreasing outwards, with radii of 
about 8\,arcsec ($\sim0.1$\,pc) for [O{\sc iii}], 15\,arcsec ($\sim0.17$\,pc) for 
H$\alpha$, and 20\,arcsec  ($\sim0.22$\,pc) for [S{\sc ii}], respectively.
The ionization gradient centred in LS\,1883 suggests that this star is the 
main ionizing source. The offset
of this source, both in the optical and the IR can be explained as
follows. LS\,1883 is producing a blister-like H{\sc ii} region 
opening the cavity to the south-west, and is carving its natal molecular cloud
to the north-east where most of the IR sources of the cluster are located.
\cite{1984A+A...138..380W} detected a splitting in the H$\alpha$ 
nebular profile centred in this O9.5\,V star of about 25 km\,s$^{-1}$, 
suggesting an expanding shell with a velocity of about 12 km\,s$^{-1}$.

Figures \ref{cumulos}C and \ref{cumulos}D show an enlargement of the H band image 
overlapped with the contours of the MSX band A, and the ratio between MSX bands 
A and C, respectively.
We must note that the maximum emission detected on band A 
(which traces PAH dust emission) lies roughly 
at the apparent projected centre of the IR star cluster.
The ratio between MSX A and C bands shows that the {\em warmest dust} 
is located at the position of the very intriguing emission-line star SS73\,24
(labeled with letter A in Figure 1B).
\cite{2004A+A...418..563R} show similar contour maps for G287.84--0.82 but
their MSX contours are shifted 11\,arcsec in declination, and they misplace
the peak, matching it with LS\,1883. The astrometric solution used for our images
is supported by the perfect match between the MSX A/C image
and the coordinates of SS73\,24 in figure \ref{cumulos}D. The link
between SS73\,24 and the IR cluster will be further discussed below.

\begin{figure}
\centering
\vspace{6.2cm}
 \includegraphics{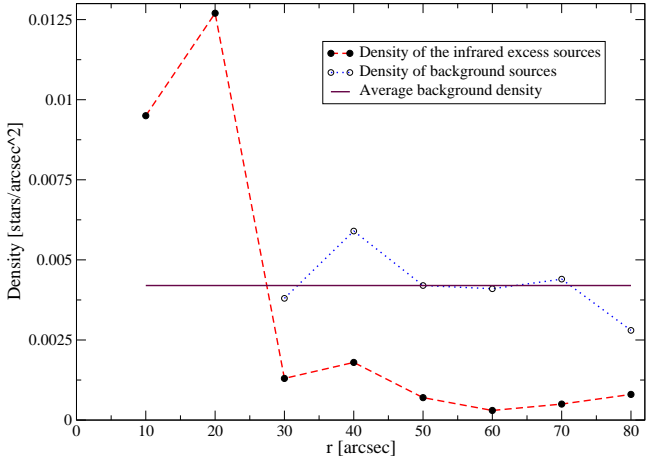}
\caption[]{Projected stellar density around 
$\alpha_{\rm 2000}$=10h\,45m\,53.6s, 
$\delta_{\rm 2000}$=-59$^{\circ}$\,57\arcmin\,02\arcsec\,.
Dashed and dotted lines refer to the density of IR excess 
and background (i.e.\ with and without IR excess) sources, respectively. 
The average background density seems to have a constant value of about 0.0042 
stars\,arcsec$^{-2}$ and is represented by a solid line.}
\label{densi}
\end{figure}  

As was noted earlier, \cite{2003A&A...400..533D} detected a
concentration of IR sources around LS\,1883 in the 2MASS images. 
The analysis of IR images are advantageous because they reduce the effect 
of the high extinction present in the zone, and also because
the controversial variation of $R_V$ in the Carina nebula does not 
affect IR extinction. This means that, in the IR, we need to consider only one 
extinction law (see Mathis 1990)\nocite{1990ARA&A..28...37M} and we can safely
neglect the dependence of the IR absorption coefficients with $R_V$. 
Bearing in mind possible projection effects, we analysed the spatial 
distribution of stars near LS\,1883.
To quantify this, we have studied the surface density of the IR excess
sources with respect to the background (Figure \ref{densi}), which was calculated
between 30\,arcsec and 80\,arcsec from the centre of the IR cluster with 
a bin size of 10\,arcsec.
The plot also includes the density of background sources, 
which shows a very flat profile, with an average value of 0.0042 
stars\,arcsec$^{-2}$.
This value is about one order of magnitude lower than the estimated density 
for the IR excess sources inside the circle of 22.5\,arcsec radius.
Another fact that supports the presence of a concentration of IR sources is 
that the density of these objects outside the 30\,arcsec radius drops roughly 
five times below the average background density.

\begin{table*}
\centering
\caption[]{2MASS photometry for the IR star cluster candidates. 
Columns 1 and 2 indicate the J2000 stellar coordinates. The
2MASS J\,H\,K$_{\rm s}$ magnitudes and their corresponding errors
are listed in columns 3-8. IR colours (H-K$_{\rm s}$) and (J-H) and 
their errors are also included in columns 9-12. Column 13 shows the 
identification number, according to Figure \ref{cumulos}B.}
\begin{tabular} {l l c c c c c c c c c c l}
\hline
$\alpha(J 2000.0)$ & $\delta(J 2000.0)$& J & $\sigma_{\rm J}$& H & $\sigma_{\rm H}$ & K$_{\rm s}$ & $\sigma_{\rm K_s}$& 
(H-K$_{\rm s}$)& $\sigma_{\rm (H-K_s)}$ & (J-H) & $\sigma_{\rm (J-H)}$ & Id. \\
\hline
161.460265 &-59.951054 & 14.22 & 0.02 & 13.18 & 0.06 & 12.55 & 0.05 & 0.63  &  0.07  & 1.40  &  0.06 &  1 \\
161.464906 &-59.952431 & 16.24 & 0.24 & 14.71 & 0.18 & 13.81 & 0.16 & 0.90  &  0.24  & 1.53  &  0.30 &  2\\
161.464823 &-59.947571 & 13.42 & null & 12.76 & 0.07 & 11.77 & 0.07 & 1.00  &  0.10  & 0.66  &  null &  3\\
161.466978 &-59.945354 & 15.19 & 0.16 & 12.57 & null & 11.40 & null & 1.17  &  null  & 2.62  &  null &  4\\
161.468003 &-59.950760 & 14.04 & 0.11 & 13.10 & 0.09 & 12.41 & 0.08 & 0.69  &  0.12  & 0.95  &  0.15 &  5\\
161.470004 &-59.952248 & 14.56 & 0.16 & 13.83 & 0.15 & 12.72 & 0.13 & 1.11  &  0.20  & 0.73  &  0.22 &  6\\
161.471684 &-59.953056 & 14.34 & 0.13 & 13.67 & 0.11 & 13.02 & 0.10 & 0.65  &  0.15  & 0.66  &  0.18 &  7\\
161.471572 &-59.943806 & 15.17 & 0.08 & 12.62 & 0.06 & 11.23 & 0.05 & 1.39  &  0.07  & 2.55  &  0.10 &  8\\
161.472295 &-59.947170 & 14.67 & 0.10 & 13.42 & 0.12 & 12.41 & 0.10 & 1.01  &  0.16  & 1.28  &  0.16 &  9\\
161.473307 &-59.944599 & 15.92 & 0.15 & 14.29 & 0.12 & 12.96 & 0.10 & 1.33  &  0.16  & 1.62  &  0.20 & 10\\
161.476696 &-59.944351 & 15.00 & 0.07 & 13.36 & 0.06 & 12.16 & 0.05 & 1.20  &  0.08  & 1.64  &  0.09 & 11\\
161.478586 &-59.945732 & 14.59 & 0.06 & 12.76 & 0.07 & 11.55 & 0.06 & 1.21  &  0.09  & 1.83  &  0.09 & 12\\
161.473868 &-59.948837 & 13.33 & 0.08 & 12.07 & 0.07 & 10.99 & 0.05 & 1.15  &  0.08  & 1.25  &  0.11 & 13\\
161.480326 &-59.947094 & 15.78 & 0.15 & 13.60 & 0.11 & 12.12 & 0.09 & 1.48  &  0.14  & 2.18  &  0.18 & 14\\
161.481440 &-59.948910 & 13.90 & 0.04 & 13.41 & 0.06 & 12.71 & 0.09 & 0.71  &  0.11  & 0.49  &  0.07 & 15\\
161.478012 &-59.948570 & 14.04 & 0.10 & 12.73 & 0.10 & 11.70 & 0.08 & 1.03  &  0.12  & 1.32  &  0.14 & 16\\
\hline	   	       	 	  	     	 		   	      	 	 	 	     	      	       
161.473788 &-59.951069 & 9.93  & 0.02 & 9.66  & 0.03 & 9.44  & 0.03 & 0.22  &  0.04  & 0.27  &  0.04 & B\\
\hline
\end{tabular}
\label{sources}
\end{table*}

\subsection{2MASS Stellar photometry}

\begin{figure}
\centering
\vspace{7.1cm}
 \includegraphics{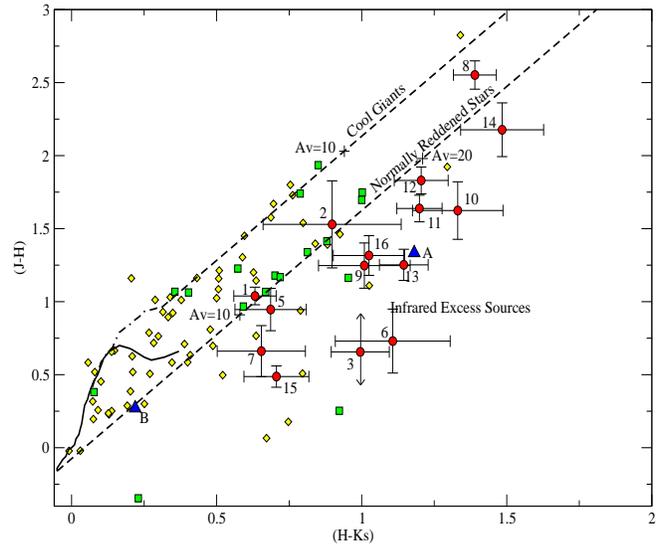}
\caption{The J\,H\,K$_{\rm s}$ C-C diagram for an 80\,arcsec
field centreed in 
$\alpha_{\rm 2000}$=10h\,45m\,53.6s, 
$\delta_{\rm 2000}=-59^{\circ}\,57'\,02.0''$. 
The unreddened main-sequence locus between O3\,V and M2\,V is indicated with 
a solid curve, while the cool giant branch is denoted with a dash-dotted line. 
The reddening track for normal O3\,V and cool giant stars are plotted as 
dashed lines, with crosses indicating A$_{\rm v}$ = 10 and 20 mag of 
(normal) visual extinction from the points where they intersect the main 
sequence and cool giant curves, respectively. 
Filled circles refer to IR cluster stars, squares to suspected foreground 
or background stars between the circles 22.5 and 40\,arcsec of radii; 
diamonds to field stars, and triangles to the luminous optical stars labelled 
with A (SS73\,24) and B (LS\,1883). Numbers label the IR sources from Table 
\ref{sources}. Error bars represent the possible values within 
the 95 per cent confidence level, and are plotted only for the IR cluster stars.}
\label{CC}
\end{figure}

\begin{figure}
\centering
\vspace{7.1cm}
 \includegraphics{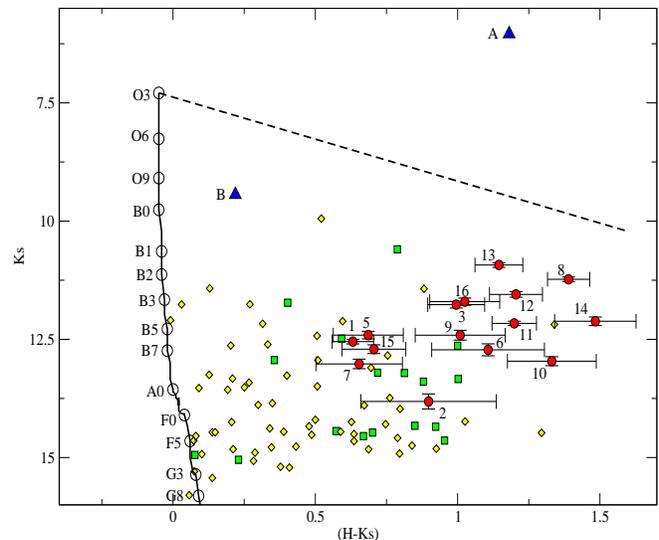}
\caption[]{The J\,H\,K$_{\rm s}$ C-M diagram for the same region 
as Figure \ref{CC}. The ZAMS between O3\,V and G8\,V is indicated with 
a solid curve and it corresponds to a distance modulus of 11.76. 
The reddening track for a ZAMS O3\,V star is plotted with a dashed line.
Labels, symbols and error bars are as in Figure \ref{CC}.}
\label{CM}
\end{figure}

To investigate the IR properties of stars in the observed field, we have
plotted J\,H\,K$_{\rm s}$ colour-colour (C-C) and colour-magnitude 
(C-M) diagrams, and compared them with the absolute IR magnitudes 
and colours from ZAMS stars (see Figures \ref{CC} and \ref{CM}). 
The main-sequence and the cool giant branch loci in the C-C diagram were 
adopted from \cite{1997ApJ...489..698H}, 
whereas the ZAMS IR magnitudes and colours for the C-M diagram were 
calculated again by us following the same procedure as 
\cite{1997ApJ...489..698H}, but including stars with later spectral types, 
reaching G8\,V. 
We have used the calibration between effective temperatures and spectral types
presented in \cite{1989AJ.....98.1305M} for O3 to B0.5, and in 
\cite{1995ApJS..101..117K} from B1 to G8, which show consistency where 
they overlap. Luminosities were obtained from the Geneva models 
\citep{1992A&AS...96..269S} as a function of these effective temperatures. 
Bolometric corrections were derived from the stellar effective temperatures
using either \cite{1989AJ.....98.1305M} or \cite{1995ApJS..101..117K}, 
accordingly. ZAMS absolute V magnitudes were therefore determined, 
and using the $(V-K)_{0}$ colours given by \cite{1983A&A...128...84K},
we calculated the ZAMS absolute K magnitudes as a function of spectral type.

Under normal circumstances, stars that appear only in the reddening band 
(between the dashed lines) in the C-C diagram (see Figure \ref{CC}), 
should be main-sequence or cool giant stars affected by different 
degrees of interstellar extinction. 
Young stellar objects (YSOs), such as Herbig Ae/Be and T\,Tauri type stars are 
frequently found to the right of the normal reddening vector of a dwarf 
early-type star because of near-IR excess emission \citep{1992ApJ...393..278L}. 
As we can see in the C-C diagram, most of the suspected IR
cluster's stars lie to the right of the reddening band, with roughly the 
same intrinsic IR excess. We have included in this C-C diagram the 
error bars estimated from error propagation of the individual magnitude errors 
as listed by the 2MASS. From these, it can be seen that the location of the IR 
cluster's sources cannot be blamed on photometric uncertainties.

Important information about the IR properties of the stars can be deduced 
from the location in the near-IR C-M diagram. 
We have assumed that the IR cluster is placed at the same distance as \ec, 
about 2300 pc \citep{1997AAS...191.3406D}, so that at such a distance, 
stars with a large (H-K$_{\rm s}$) colour are distant cool giant and 
supergiant stars, embedded luminous O\,B 
stars, or very luminous YSOs. In the latter case, this excess is an 
intrinsic signature of the source.
With these, we built the C-M diagram (Figure \ref{CM}) which 
shows that stars with IR excess form a distinct group in the plane, 
lying in the region where pre-main sequence stars are expected to be observed.
Source 4 has not been included in both diagrams due to the great 
uncertainty in the magnitude values (see Table \ref{sources}). 
However, we have taken into account this object in the estimation of the 
projected stellar density because it is a clear IR source. 
In the case of source 3 the $(J-H)$ error bar is plotted with arrows,
indicating the uncertainty in the $J$ magnitude error. However, 
as the magnitudes in the other two bands are acceptable, and so its position 
in the C-M diagram is not affected, we have included it in this diagram.

\begin{figure*}
\centering
\vspace{6.5cm}
\includegraphics{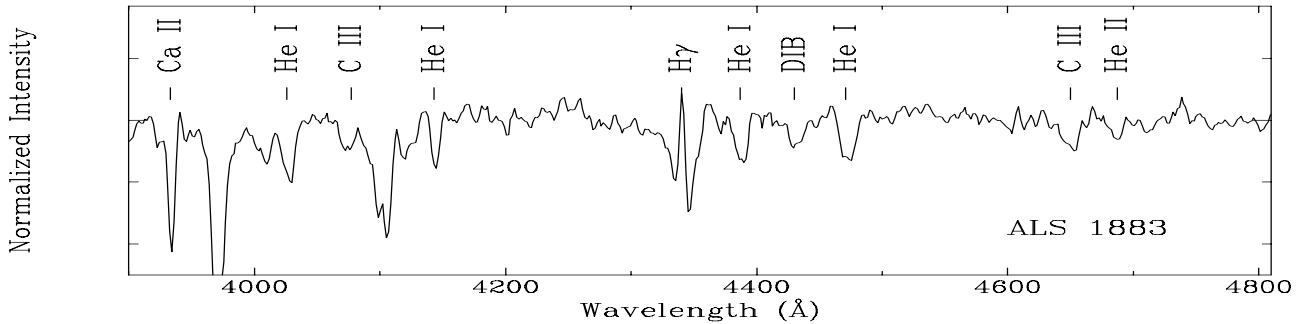}
\caption{Normalized blue spectrum of LS\,1883 obtained at CASLEO. The O9.5\,V
spectral classification is derived from the presence of He{\sc ii} 
4686\AA\,, and the barely visible He{\sc ii} 4541\AA\, absorption lines
compared with He I 4471\AA\,, and also from the moderately strong 
C{\sc iii} 4650\AA\, absorption line.}
\label{esp}
\end{figure*}

On the other hand, we can appreciate that most of the observed field 
stars in Figure \ref{cumulos}B (out to 22.5\,arcsec from the centre of the cluster)
delineate in both diagrams a typical distribution of main-sequence and cool 
giant field stars affected by a low and moderate IR extinction.

\cite{2003A&A...400..533D} estimated for this cluster 10 members and 
a radius of 45\,arcsec (0.55 pc) at a distance of 2500 pc, which means a 
density of about 15 stars pc$^{-3}$. This result is much lower than the 
densities of  10$^{2}$ to 10$^3$ expected for IR clusters with current star 
formation - values which were obtained from a study of 28 young stellar 
clusters surrounding Herbig Ae/Be stars \citep{1999A&A...342..515T}.
As we have shown above, if we assume that the IR cluster has at least 
17 resolved members and a radius of 22.5\,arcsec ($\approx$ 0.25 pc at 
a distance of 2300 pc), we can derive a lower limit for the stellar density 
of about 260 stars pc$^{-3}$, in better agreement with 
\cite{1999A&A...342..515T}. This lower limit is established by the 
resolution and sensitivity of the 2MASS images.

\subsection{LS\,1883 \& SS73\,24}

The early spectral type of the optically brightest star (LS\,1883), 
which lies inside the estimated IR cluster radius (see Figure \ref{cumulos}B), 
constrains the scenarios described by \cite{1999A&A...342..515T},
where it is expected that clusters with such high densities have associated
massive stars. Alternatively, this star could be a field star projected on the 
IR cluster sky position. From our low dispersion spectra of LS\,1883, 
we have classified it as an O9.5\,V star using the relative intensities 
of He{\sc i} 4471\AA\, and He{\sc ii} 4542\AA\, and He{\sc ii} 4686\AA\ 
absorption lines, according to the Conti (\citeyear{1973ApJ...179..181C}) 
criteria (Figure \ref{esp}).
Furthermore, the comparison with the Walborn \& Fitzpatrick 
(\citeyear{1990PASP..102..379W}) OB Stars Spectral Atlas shows that the 
spectrum of LS\,1883 
resembles that of O9.5\,V star HD\,93027, including the presence of 
C\,{\sc iii} absorption lines. Our spectral classification of LS\,1883 
is in agreement with the previous one by \cite{1984A+A...138..380W}. 
We note that even though Walsh's sky-subtracted stellar spectrum was 
contaminated with emission lines originating from the nebula, he arrived at the 
same result as us. 

The positions of LS\,1883 (labelled B in Figure \ref{cumulos}B) in both the 
C-C and C-M diagrams (Figures \ref{CC} and \ref{CM}) 
agree very well ($\Delta K_{\rm s}$\,$\approx$\,0.38) with a moderately 
reddened O9\,V star, which suggests that, within the observational errors, 
the star could belong to the cluster. In this case, it can be interpreted as a
ZAMS massive star at the beginning of the destruction of its natal 
cocoon, perhaps in one of the earliest stages of evolution compared 
with the Trapezium O stars in the Orion Nebula. The ultraviolet radiation and 
incipient stellar winds from the O9.5\,V star LS\,1883 seem to produce a 
small blister-like nebular structure, carving its parental cloud and
revealing the embedded cluster whose members probably have 
lower masses, in a similar scenario found in Knots 1 and 2 in the 30 
Doradus nebula \citep{1999AJ....117..225W}.

The second conspicuous optical star (labelled A in Figure \ref{cumulos}B) is
SS73\,24 (WRAY 15-642), a Be!\,pec star \citep{1973ApJ...185..899S}.
The spectrum was briefly described by \cite{2003A&A...397..927P}, who have shown 
the presence of Balmer lines, He{\sc i} 5876 \AA\, and some iron lines in
emission (e.g. Fe{\sc ii}, [Fe{\sc ii}]). Moreover, \cite{1999ApJ...526..854G}
have found that the spectrum resembles that of Hen\,401, a protoplanetary
nebula. Our spectra of SS73\,24 look very similar to those presented
by \cite{2003A&A...397..927P}. The very careful background subtraction
shows that no forbidden emission lines of [O{\sc i}] 6300-6363 \AA\, and 
[S{\sc ii}] 6716-6731 \AA\, are seen, but S{\sc ii} 6347-6350 \AA\,, 
He{\sc i} 6678 \AA\,, and some additional Fe{\sc ii} emission lines are 
present.
Furthermore, the position of the star in the C-C diagram 
(Figure \ref{CC}), shows that the IR excess for this star 
is consistent with most of the IR cluster stars, suggesting that this star 
could be linked with the IR cluster. However, the position of this star in 
the C-M diagram (Figure \ref{CM}) clearly shows that it is brighter than the 
rest of the IR cluster's stars. It also lies above the O3\,V reddening vector, 
which could be interpreted as due to a shorter distance to this object 
(and thus unrelated to the IR cluster) or due to the presence of strong 
intrinsic IR continuum emission. In this way, as we note above, SS73\,24 
is detected in the MSX band A image as an extension of the emission to the 
east of G287.84--0.82 (Figure \ref{cumulos}C). Also, this star presents an
MSX band A to C ratio, which indicates warm dust emission directly 
associated with it. The inspection of our optical narrow band images 
does not reveal any nebular structure associated with SS73\,24. The absence of
forbidden emission lines of [S{\sc ii}] 6716-6731\AA\ suggests that the star 
is surrounded by a very dense and 
unresolved compact dusty envelope. Peculiar spectra like SS73\,24 are
also found in active interacting binaries of moderately long periods which
exhibit strong mass loss, such as W Serpentis \citep{1993A&A...269..390B},
or AS\,78 and MWC\,657 \citep{2000A&AS..147....5M}. SS73\,24 deserves 
 more special attention (both spectroscopically and photometrically), 
in order to understand its nature, and whether or not it is related to the cluster.

\section{Summary}

We have derived some of the optical and near-IR properties of G287.84--0.82 
confirming that it is an IR star cluster, embedded in the head of a dusty
pillar in the southern part of the Carina nebula (NGC\,3372). Also, we
have established some important properties of the brightest optical sources
in the field of the cluster.
These can be summarized as follows.
\begin{enumerate}
\item The surface density of IR excess sources located in the IR
cluster is one order of magnitude greater than the background density.
\item The density of IR excess sources outside the 30\,arcsec radius 
is roughly five times lower than the background density.
\item The cluster has at least 17 members, and its parameters, radius and 
stellar density, are in good agreement with high to intermediate mass star 
formation scenarios.
\item C-C and C-M diagrams show roughly the same IR excess as most of 
the IR cluster member candidates.
\item These IR excesses suggest that these objects could be 
intermediate-mass pre-main sequence stars. 
\item LS\,1883 is located near to the IR cluster core. 
We confirm the O9.5\,V spectral type for this star based on the
presence of He\,{\sc ii} absorption lines. This spectral classification
agrees with the position in the C-C and C-M diagrams for a moderately 
reddened O9\,V star. LS\,1883 seems to be the
first ZAMS star emerging from the cluster, and it could be producing a
blister-like H{\sc ii} region. Optical narrow band images reveal an ionizing
gradient in the nebula surrounding the star.
\item The positions of SS73\,24 in the C-C and C-M diagrams suggest that 
this star has a strong intrinsic near-IR continuum emission, although 
we cannot establish any relationship between SS73\,24 and the IR cluster. 
\item The spectral characteristics of SS73\,24 suggest that this star could
be related to a strong mass-loss stage in an interacting binary.
\item The mid-IR MSX images show emission from the IR cluster and from
warm dust associated with SS73\,24. 
\end{enumerate}

Some of these facts are consistent with the current star formation scenarios 
associated with highly embedded IR star clusters, where the most massive
component is an O-type star. 
In order to be able to reveal the true nature of each individual IR cluster
member, spectroscopic observations and images with higher spatial resolution
 in the near-IR and mid-IR wavelength range are needed.

\vskip 0.5cm

{\it Acknowledgments} We are very grateful to Michael Taylor and Silvia Giovagnoli for 
their aid in the clarification of some passages of the text. We specially want to thank
Elena Terlevich, who read this paper and suggested to us a better way to 
present different aspects of this work. Substantial improvements to the
paper resulted from a careful revision and suggestion from an anonymous
referee.
This research made use of data products from the Midcourse Space 
Experiment. Processing of the data was funded by the Ballistic 
Missile Defense Organization with additional support from NASA 
Office of Space Science. This research has also made use of 
data products from the Two Micron All Sky Survey, which is a joint project
of the University of Massachusetts and the IR Processing and Analysis 
Centre, funded by the NASA and the NSF. 
We want to thank the director and staff of CTIO for the use of their facilities 
and kind hospitality during the observing run.
We acknowledge the use at CASLEO of the CCD and data 
acquisition system supported under US NSF grant AST-90-15827 to R.M. Rich.
Also, we thank Julia Arias for providing us with the low resolution 
spectra of LS\,1883 and SS73\,24.

\bibliographystyle{mn2e}
\bibliography{ir}

\end{document}